\documentclass[aps,prl,letterpaper,twocolumn,showpacs,superscriptaddress,floatfix,longbibliography]{revtex4-1}

\usepackage[english]{babel} \usepackage{amsmath} \usepackage{color}
\usepackage{epsfig} \usepackage{graphicx} \usepackage{bm}
\usepackage{mathtools}

\usepackage{times,amsmath,amssymb} \usepackage{amsfonts}
\usepackage{amssymb}

\usepackage[colorlinks,urlcolor=blue,bookmarks=false,hypertexnames=true]{hyperref}

\newcommand{\tr}{\mathop{\mathrm{tr}}\limits}

\newcommand{\trzero}{\mathop{\mathrm{tr}_{\mathcal{H}_0}}\limits}

\global\long\def\bra#1{\langle #1 |} \global\long\def\ket#1{| #1
  \rangle } \global\long\def\braket#1#2{\langle #1|#2 \rangle }

\global\long\def\al{\alpha} \global\long\def\be{\beta}
\global\long\def\ga{\gamma} \global\long\def\de{\delta}
\global\long\def\De{\Delta} \global\long\def\Ga{\Gamma}
\global\long\def\th{\theta}

\global\long\def\la{\lambda} 
\global\long\def\si{\sigma} \global\long\def\vfi{\varphi}

\global\long\def\tiD{\tilde{\cal D}}

\global\long\def\cL0{{\cal L}_0}

\global\long\def\tde{\tilde{\delta}}
\global\long\def\tga{\tilde{\gamma}}
\global\long\def\teps{\tilde{\epsilon}}
\global\long\def\tsi{\tilde{\sigma}}

\global\long\def\eps{\epsilon}

\global\long\def\bege{\begin{equation}}
  \global\long\def\ende{\end{equation}}

\global\long\def\begal{\begin{align}}
  \global\long\def\endal{\end{align}}

\allowdisplaybreaks

\def\RE{\mathop{\rm Re}} %
\definecolor{myred}{RGB}{168,5,14}
\definecolor{myblue}{RGB}{13,13,255}
\definecolor{editorcolor}{RGB}{220,13,255}
\definecolor{mygreen}{RGB}{20,150,20}

\begin{document}

\title{Full Spectrum of the Liouvillian of Open Dissipative Quantum
  Systems in the Zeno Limit}

\author{Vladislav Popkov} \affiliation{Department of Physics,
  University of Wuppertal, Gaussstra\ss e 20, 42119 Wuppertal,
  Germany}

\author{Carlo Presilla} \affiliation{Dipartimento di Fisica, Sapienza
  Universit\`a di Roma, Piazzale Aldo Moro 2, Roma 00185, Italy}
\affiliation{Istituto Nazionale di Fisica Nucleare, Sezione di Roma 1,
  Roma 00185, Italy}

\date{\today}

\begin{abstract}
  We consider an open quantum system with dissipation, described by a
  Lindblad Master equation (LME).  For dissipation locally acting and
  sufficiently strong, a separation of the relaxation timescales
  occurs, which, in terms of the eigenvalues of the Liouvillian,
  implies a grouping of the latter in distinct vertical stripes in the
  complex plane at positions determined by the eigenvalues of the
  dissipator.  We derive effective LME equations describing the modes
  within each stripe separately, and solve them perturbatively,
  obtaining for the full set of eigenvalues and eigenstates of the
  Liouvillian explicit expressions correct at order $1/\Gamma$
  included, where $\Ga$ is the strength of the dissipation.  As an
  example, we apply our general results to quantum $XYZ$ spin chains
  coupled, at one boundary, to a dissipative bath of polarization.
\end{abstract}

\maketitle

Recently a great deal of analytic progress has been made in the theory
of open quantum systems and their steady-state exact solutions.  Much
less is known about the full spectrum of the Liouvillian [the
Lindbladian, more precisely, if the open quantum system is described,
as very often happens, by a Lindblad master equation (LME)].  Just to
say, the knowledge of this spectrum is essential to predict the
finite-time evolution of dissipative systems, as of interest in fields
ranging from quantum computing~\cite{QuantumComputing} to quantum
biology~\cite{QuantumBiology}.  The problem basically remains
intractable, except via hard computational
methods~\cite{Briegel.Englert,Barnett.Stenholm,Rocca2008,Torres.Betzholz.Bienert}.

The existing literature regarding the Liouvillian general properties
focuses on an analysis of asymptotic time regime $t \to \infty$, i.e.,
putting emphasis on the existence of a decoherence-free subspace and
the asymptotic leakage out of
it~\cite{Marcuzzi.etal,GeometryAndResponse,Shpielberg}. Within such an
approach, however, a substantial part of information about the
Liouvillian spectrum is lost.

Exceptionally, under special conditions imposed on the Lindblad
operators and the Hamiltonian, the Liouvillian spectrum can be related
to the spectrum of auxiliary non-Hermitian operators. However, even in
this case, the complete set of eigenstates is out of
reach~\cite{MEP,Kurlov,BBMJ,NKM}.

In the present communication, in contrast, we show how to obtain the
complete set of eigenvalues and eigenstates of the Liouvillian,
provided that the dissipation is sufficiently strong with respect to
the coherent part of the evolution, in the so-called quantum Zeno
regime~\cite{Misra1977,POT1996,ZenoStaticsExperimentalReview,ZenoPascazio2020}.
For this setup to be nontrivial, dissipation must act only on a part
of the degrees of freedom.

As we will see, in the limit of strong dissipation acting on a part of
degrees of freedom, the behavior of an open quantum system simplifies
and the full Liouvillian can be block diagonalized.

We provide a general procedure to obtain the full set of eigenvalues
and eigenstates by means of a perturbative approach in terms of the
solution of a linear problem for the dissipation-projected Hamiltonian
\cite{2018ZenoDynamics,2014Venuti}, and other related Hamiltonians
acting in a reduced Hilbert space.  As an example, we comprehensively
discuss the case of general open $XYZ$ spin chains with arbitrary spin
states targeted at one of the boundaries by the strong interaction
with dissipative environments.

\textit{General theory.}---We Considery an open quantum system with
finite Hilbert space $\mathcal{H}$ and dissipation acting only on a
part of its degrees of freedom, namely, those associated to the
subspace $\mathcal{H}_0 \subset \mathcal{H}$. Denoting by
$\mathcal{H}_1$ the dissipation-free subspace, we have
$\mathcal{H}=\mathcal{H}_0 \otimes \mathcal{H}_1$ with
$\dim \mathcal{H}_0 = d_0$, $\dim \mathcal{H}_1 = d_1$ and
$d_0 d_1=d=\dim \mathcal{H}$.  The evolution of the reduced density
matrix operator of the systems,
$d\rho(\tau)/d\tau = \mathcal{L}[\rho(\tau)]$, is determined by the
Liouvillian
\begin{align}
  \label{L}
  \mathcal{L}[\cdot] = - i [H,\cdot] + \Gamma \mathcal{D}[\cdot],
\end{align}
where $H$ is the Hamiltonian of the system, $\mathcal{D}[\cdot]$ a
Lindblad dissipator of standard form and $\Gamma$ the strength of the
dissipation. The use of a Markovian Lindblad dynamics for large
dissipation is justified for reservoirs with very short correlation
times~\cite{Gorini}. Note that we work in units of $\hbar=1$, i.e.,
$\tau=t_\mathrm{ph}/\hbar$ and $\Gamma=\Gamma_\mathrm{ph}\hbar$, where
$t_\mathrm{ph}$ and $\Gamma_\mathrm{ph}$ are the physical time and
dissipation strength.

In Ref.~\cite{2018ZenoDynamics} it has been shown that in the Zeno
limit $\Gamma\to\infty$ the dynamics~(\ref{L}) is still reduced to a
new Lindblad equation written in terms of a renormalized Hamiltonian
and an effective dissipator.  More precisely, for times
$\tau \gg 1/\Gamma$ and with an error $O(1/\Gamma^2)$ we have
$\rho(\tau)=\psi_0 \otimes R_0(\tau)$, where $\psi_0\in\mathcal{H}_0$
is the dissipator kernel, $\mathcal{D}[\psi_0]=0$, (assumed to be
unique) and $R_0(\tau)\in\mathcal{H}_1$ satisfies
\begin{align}
  \frac{d R_0(\tau)}{d\tau} =
  - i [h_D+ \tilde{H}_a/\Ga ,R_0(\tau)]
  + \frac{1}{\Ga} \tiD[R_0(\tau)].
  \label{LMEeffective0}
\end{align}
The effective Hamiltonian $\tilde{H}=h_D+ \tilde{H}_a/\Ga$ is the sum
of the dissipation-projected Hamiltonian,
$h_D = \trzero((\psi_0\otimes I_{\mathcal{H}_1}) H)$, and a Lamb shift
correction $\tilde{H}_a$.  With $\trzero$ we indicate the trace in the
subspace $\mathcal{H}_0$.  Note that both $\tilde{H}_a$ and the
effective dissipator $\tiD[\cdot]$ act in the sole subspace
$\mathcal{H}_1$.  Explicit expressions of $\tilde{H}_a$ and
$\tiD[\cdot]$ are given in [6] and, for convenience, reported in
Supplemental Material~\cite{SM}.

Equation (\ref{LMEeffective0}) provides a complete information about
$R_0$, the dissipation-free component of the density matrix $\rho$.
The full density matrix has, however, an expansion of the form
$\rho(\tau)= \sum_{k} \psi_k \otimes R_k(\tau)$, where $\psi_k$ are
the eigenstates of the original dissipator $\mathcal{D}$ (which we
assume diagonalizable),
\begin{align}
  \mathcal{D}[\psi_k] = c_k \psi_k.
  \label{EqCalDeig}
\end{align}
The complex eigenvalues $c_k$ always have a nonpositive real part and
one of them is 0, conventionally, $c_0=0$.  When $\Ga$ is large, all
the components $k>0$ of the density matrix lying outside the
dissipation-free subspace can be shown to scale as $1/\Ga$, namely,
$||R_k(\tau)|| = O(1/\Ga)$ for $\tau > O(1)$, see \cite{2014Venuti}.

The spectrum of the effective Liouvillian $\tilde{\mathcal{L}}$
associated to Eq.~(\ref{LMEeffective0}) gives only a part of the full
Liouvillian spectrum, namely, $d_1^2$ eigenvalues out of
$d^2 = (d_0 d_1)^2$.  The remaining $d^2-d_1^2$ eigenvalues of the
Liouvillian $\mathcal{L}$ originate from the components $R_k$ with
$k>0$ in the expansion of the full density matrix.

In \cite{2018ZenoDynamics} it has been shown how to obtain, in the
Zeno limit, the nonequilibrium steady state, i.e., the eigenstate of
$\mathcal{L}$ corresponding to the eigenvalue $0$.  Here, we derive
explicit formulas for \textit{all} the eigenvalues and eigenstates of
$\mathcal{L}$ \textit{near} the Zeno limit, up to order $1/\Gamma$
included.  Explicitly, we will first obtain equations analogous to
Eq.~(\ref{LMEeffective0}) for all the components $R_k(\tau)$ of the
density matrix, and then show how to use these equations to derive
eigenvalues and eigenstates of $\mathcal{L}$.

In order to formulate our main statement, note that the dissipator
eigenstates $\{\psi_k\}$ of Eq.~(\ref{EqCalDeig}) form a basis in
$\mathcal{H}_0$. Let $\{\vfi_k\}$ be a biorthogonal basis in
$\mathcal{H}_0$ satisfying $\tr(\psi_k \vfi_n)= \de_{kn}$.  The
decompositions of the Hamiltonian $H$ and of the density matrix
$\rho(\tau)$ in the bases $\{\vfi_k\}$ and $\{\psi_k\}$ are,
respectively,
\begin{align}
  &H=  \sum_{m} (\vfi_m^\dagger \otimes g_m^\dagger)
    = \sum_{m} (\vfi_m \otimes g_m),
    \label{Hdecomp}
  \\
  &\qquad g_m= \trzero (( \psi_m \otimes I_{\mathcal{H}_1})H),
    \label{gk}
  \\
  &\rho(\tau) = \sum_{k} \psi_k \otimes R_k(\tau),
    \label{rhodecomp}
  \\
  &\qquad R_k(\tau)=  \trzero (( \vfi_k \otimes I_{\mathcal{H}_1})
    \rho(\tau)).
    \label{Rk}
\end{align}
 
\textit{Statement.}---The component $R_k$ corresponding to a nonzero
dissipator eigenvalue $c_k$ with degeneracy $\mathrm{deg}$, near the
Zeno limit satisfy
\begin{align}
  \label{statement}
  \frac{d R_k}{d \tau}
  =&\
     \Ga c_k  R_k + i \sum_{s:c_s=c_k}
     \left( U_{k,s} R_s - R_s W_{k,s}\right)
     \nonumber\\
   &
     +\frac{1}{\Ga} 
     \sum_{z>0}
     \sum_{m>0}
     \sum_{n:c_n\neq c_k}
     \sum_{s:c_s=c_k}
     \frac{1}{c_n-c_k}
     \nonumber\\
   &\qquad\times
     \Big( -\ga^{n,s,k}_{m,z} g_m R_s g_z^\dagger
     +\varepsilon^{n,s,k}_{z,m} g_z^\dagger g_m R_s
     \nonumber\\
   &\qquad\qquad
     +\de^{n,s,k}_{z,m} R_s g_z^\dagger g_m \Big) + O(1/\Gamma^2),
\end{align}
where $U_{k,s}$ and $W_{k,s}$ are operators in $\mathcal{H}_1$ given
by
\begin{align}
  U_{k,s}=  \sum_{n}   B_{n,s,k} g_n^\dagger,
  \quad
  W_{k,s}=  \sum_{n}   A_{n,s,k} g_n^\dagger
  \label{DefUW}
\end{align}
and $\ga^{n,s,k}_{m,z}$, $\varepsilon^{n,s,k}_{m,z}$ and
$\de^{n,s,k}_{m,z}$ are the coefficients
\begin{align}
  &\ga^{n,s,k}_{m,z}= C_{m,s,n} A_{z,n,k} +A_{z,s,n}  C_{m,n,k},
    \label{EqGa}\\
  &\eps^{n,s,k}_{z,m}= C_{m,s,n} B_{z,n,k},  \label{EqEps}\\
  &\de^{n,s,k}_{z,m}=A_{z,s,n} C_{k,n,m}, \label{EqDe} 
\end{align}
with
\begin{align}
  &A_{m,k,n} = \tr(\vfi_n \psi_k \vfi_m^\dagger), \label{DefAcoeff}\\
  &B_{m,k,n} = \tr(\vfi_n \vfi_m^\dagger \psi_k), \label{DefBcoeff}\\
  &C_{m,k,n} = \tr(\vfi_n \vfi_m \psi_k). \label{DefCcoeff}
\end{align} 
Note that the above coefficients are related to the dissipator via its
eigenstates $\{\psi_k\}$ and the associated biorthogonal basis
$\{\varphi_k\}$.  For a nondegenerate eigenvalue, $\mathrm{deg}=1$,
only the simplified operators $U_{k,k}=U_{k}$ and $W_{k,k}=W_{k}$
appear in Eq.~(\ref{statement}), where
\begin{align}
  U_{k} = g_0 + \sum_{n>0} B_{n,k,k} g_n^\dagger,
  \quad
  W_{k}  = g_0 + \sum_{n>0} A_{n,k,k} g_n^\dagger.
  \label{DefUWnd}
\end{align}
Equation~(\ref{statement}) applies also in the presence of more
degenerate eigenvalues.
 
The above statement follows from a perturbative Dyson expansion with
respect to the small parameter $1/\Ga$ of the Liouvillian equation for
$\rho(t)$, where $t$ is the scaled time $t=\Ga\tau$.  With this
scaling, we have $d\rho(t)/dt=\cL0[\rho(t)]+K[\rho(t)]$, where
$\cL0[\cdot]=\mathcal{D}[\cdot]$ and
$K[\cdot] = -(i/\Gamma) [H,\cdot]$.  The corresponding exact
propagator, namely, $\exp((\cL0+K)t)$, can be expanded in a Dyson
series with respect to the perturbation $K$. Keeping the expansion
terms up to order $K^2$ included and coming back to the time $\tau$,
after some algebra, we get Eq.~(\ref{statement}). Full details of the
proof are given in Supplemental Material~\cite{SM}.

\textit{Eigenvalues and eigenvectors of $\mathcal{L}$.}---By finding
the normal modes of the linear problem (\ref{statement}) for each
index $k>0$, as well as of the linear problem (\ref{LMEeffective0})
for $k=0$, we obtain all the eigenvalues of the Liouvillian
$\mathcal{L}$.  Let $\la_{k,\al,\be}$ be the set of the eigenvalues of
$\mathcal{L}$ corresponding to $c_k$.  First consider a nondegenerate
$c_k$.  In the limit $\Ga \rightarrow \infty$, the $O(1/\Ga)$
contributions in Eq.~(\ref{statement}) can be neglected and, expanding
$R_k(\tau) = \sum_{\al,\be} c_{\al \be}(\tau) \ket{\al} \bra{\tilde
  \be}$, where $\ket{\al} $ are the right eigenvectors of $U_k$ with
eigenvalues $u_\al$ and $\bra{\tilde\be} $ are the left eigenvectors
of $W_k$ with eigenvalues $w_\be$, we find
$dc_{\al \be}(\tau)/d\tau = \la_{k,\al,\be} c_{\al \be}(\tau) = (c_k
\Ga + i (u_\al- w_\be)) c_{\al \be}(\tau)$. This implies
\begin{align}
  &\la_{k,\al,\be} = c_k \Ga + i (u_\al- w_\be)+ O(1/\Ga),
    \label{EigenvaluesRk}
\end{align}
with corresponding eigenvectors
$\psi_{k,\al,\be}=\psi_k\otimes \ket{\al} \bra{\tilde\be}$.  Note
that, even if not explicitly indicated, the eigenvalues $w_\al$ and
$u_\be$ depend, as the corresponding eigenvectors do, on the index
$k$.

The $1/\Ga$ corrections to the eigenvalues (\ref{EigenvaluesRk}) are
then found by a standard perturbative formula,
$\de \la_{k,\al,\be}= \bra{R_{\al \be}} \hat{V}_k \ket{R_{\al \be}}$,
where $\hat{V}_k$ is the vectorized form of the $O(1/\Ga)$ term in the
superoperator of Eq.~(\ref{statement}) and $\ket{R_{\al\be}}$ is an
eigencomponent of the reduced density matrix
$\ket{R_{\al\be}}=\ket{\al} \otimes \ket{\tilde\be}^*$.  Every
perturbative term of type $Q \ket{x}\bra{y}P $ in (\ref{statement})
gives a contribution
$\tr(Q \ket{x}\bra{y}P \ket{y}\bra{x})= \bra{x}Q \ket{x}\bra{y}P
\ket{y}$ to the eigenvalue correction.  Explicitly, we obtain
\begin{align}
  &\de \la_{k,\al,\be}
    =\  \frac{1}{\Ga}
    \sum_{z>0}\sum_{m>0}  \sum_{n:c_n \neq c_k} \frac{1}{c_n-c_k}
    \Big(
    -\ga^{n,n,k}_{m,z}   \bra{\al} g_m \ket{\al}
    \nonumber\\
  &\times
    \bra{\tilde \be} g_z^\dagger  \ket{\tilde \be}
    +
    \varepsilon^{n,n,k}_{z,m}  \bra{\al} g_z^\dagger g_m \ket{\al}   + 
    \de^{n,n,k}_{z,m}  \bra{\tilde \be}   g_z^\dagger g_m  \ket{\tilde \be}
    \Big).
    \label{CorrectionEigenvaluesRk}
\end{align}
The $O(1/\Gamma)$ corrections to the respective eigenstates
$\psi_k\otimes \ket{\al} \bra{\tilde\be}$ are also given by standard
first-order perturbative formulas~\cite{Landau}.

Of course, the above $1/\Gamma$ correction is valid if the eigenvalues
(\ref{EigenvaluesRk}) are nondegenerate. In the case of
$\la_{k,\al,\be}$ degenerate, a different, although still standard,
procedure must be undertaken (diagonalization within the subspace of
degeneration) to obtain the $1/\Ga$ corrections.  Explicit expressions
will be given for the case study considered below.

The case of a degenerate dissipator eigenvalue $c_k$ can be tackled in
a similar way.

\textit{A case study: The $XYZ$ spin chain.}---We illustrate the above
results on a Heisenberg spin chain with $N+1$ sites, the first one
being in contact with a strongly dissipative environment. The coherent
part of the evolution is given by the standard $XYZ$ Hamiltonian
$H= \sum_{j=0}^{N-1} \sum_{\alpha=x,y,z} \sigma_j^\alpha J_\alpha
\sigma_{j+1}^\alpha$, $\sigma_j^\alpha$ being the $\alpha$-th Pauli
matrix acting at site $j$, whereas dissipation acts locally on site
$0$ and targets an arbitrary, pure or mixed, single spin state
$\rho_0$ at this site~\cite{Prosen2011}.  The evolution of the the
density matrix $\rho(\tau)$ of the full chain is determined by a LME
with Liouvillian as in Eq.~(\ref{L}).  The Lindblad dissipator acting
on spin $1$ is the sum of two terms,
$\mathcal{D} = ((1+\mu)/2) \mathcal{D}_1 + ((1-\mu)/2) \mathcal{D}_2$,
\begin{align}
  \mathcal{D}_\al [\rho] =
  L_\al  \rho L_\al ^\dagger - \frac{1}{2} L_\al ^\dagger L_\al 
  \rho - \frac{1}{2}\rho L_\al ^\dagger L_\al, \quad
  \al=1,2,
  \label{DefLMEDissipator}
\end{align}
where $L_1=\ket{s(\th,\vfi)}\bra {s^\perp(\th,\vfi)}$ and
$L_2= L_1^\mathrm{T}$, with
$\ket{s(\th,\vfi)}=\cos(\theta/2)e^{-i\varphi/2}
\ket{{\uparrow}}+\sin(\theta/2)e^{i\varphi/2} \ket{{\downarrow}}$ and
$\braket{s(\th,\vfi)}{s^\perp(\th,\vfi)}=0$.  This dissipator targets
the polarization $\mu \vec{n}_0$ on site $0$, where $\vec{n}_0$ is the
unit vector
$\vec{n}_0=(\sin \th \cos \vfi,\sin \th \sin \vfi,\cos \th)$.  The
uniqueness of the nonequilibrium stationary state (NESS) can be proven
using Evans criterion \cite{Evans}.

\begin{figure}
  \includegraphics[width=\columnwidth]{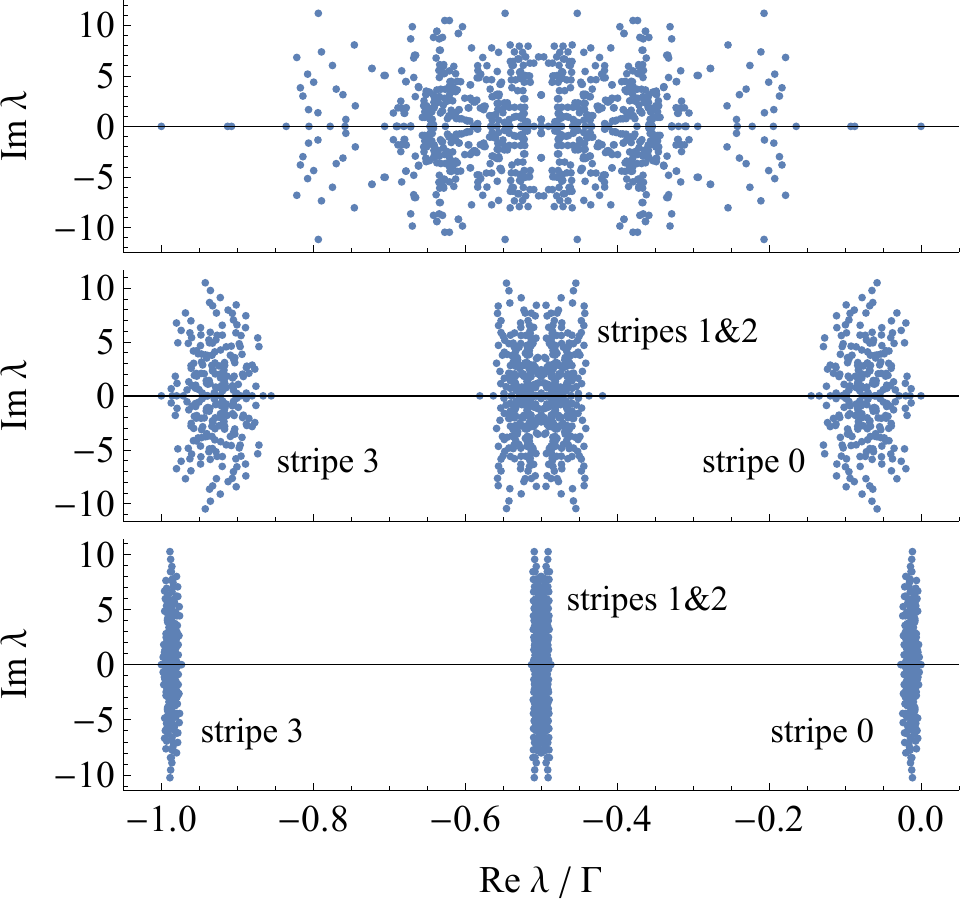}
  \centering
  \caption{(color online) Exact complex eigenvalues of the Liouvillian
    evaluated numerically for $\Ga=0.5,\ 8,\ 20$ (from top to bottom).
    Increasing $\Ga$, eigenvalues arrange in stripes whose number
    equals that of the eigenvalues of the dissipator.  For $\Ga$
    large, the width of the stripes scales as $1/\Ga$ while their
    height remains constant; the distance between the stripes scales
    as $\Gamma$. Parameters: $N=4, \vec{J}=(1,1,-0.6058)$,
    $\varphi=0$, $\theta=\pi/2$, $\mu=1$. }
  \label{Fig-LiuAllGamma05-20}
  \centering
\end{figure}

\textit{Striped structure of spectrum.} The distribution of the
Liouvillian eigenvalues manifestly depends on the strength of
dissipation $\Gamma$ in Eq.~(\ref{L}).  For medium dissipation
strengths, comparable with the exchange integral in the model, the
eigenvalues are scattered seemingly randomly, see
Fig.~\ref{Fig-LiuAllGamma05-20} (top).  For large $\Ga$, they are
arranged in distinct stripes, see Fig.~\ref{Fig-LiuAllGamma05-20}
(bottom). The stripelike structure stems from the properties of the
dissipator in the LME. In fact, the eigenvalue problem
(\ref{EqCalDeig}) of the locally acting dissipator $\mathcal{D}$ can
be easily solved~\cite{2018ZenoDynamics}, yielding
\begin{align}
  \begin{array}{ll}
    c_0=0,\qquad
    &\psi_0= \frac{1+\mu}{2} \ket{s}\bra{s} + \frac{1-\mu}{2}
      \ket{s^\perp} \bra{s^\perp},
    \\
    c_1=-\frac{1}{2},\qquad
    &\psi_1 =\ket{s}\bra{s^\perp},
    \\
    c_2=-\frac{1}{2},\qquad
    &\psi_2=\ket{s^\perp}\bra{s},
    \\
    c_3=-1,\qquad
    &\psi_3 =\ket{s}\bra{s} - \ket{s^\perp} \bra{s^\perp}.
  \end{array}
      \label{DissEig}
\end{align}
where $\ket{s} \equiv \ket{s(\theta,\vfi)}$.  The respective
biorthogonal basis $\{\vfi_k\}$ is given by
\begin{align}
  \begin{array}{l}
    \vfi_0 = I_\mathcal{H},
    \qquad
    \vfi_1 = \psi_2,
    \qquad
    \vfi_2 = \psi_1,
    \\
    \vfi_3 = \frac{1-\mu}{2} \ket{s}\bra{s} - \frac{1+\mu}{2} \ket{s^\perp}
    \bra{s^\perp}.
  \end{array}
  \label{biorthobasis}
\end{align}
Neglecting the coherent part provided by the Hamiltonian $H$, the
Liouvillian $\mathcal{L}$ of Eq.~(\ref{L}) would have the eigenvalues
$\Gamma c_k$, $k=0,\dots,3$, each eigenvalue having a degeneracy
$2^{2N}$ due to the inclusion of the $N$ extra spins.  Adding $H$ acts
as a perturbation (the small parameter being $1/\Ga$), which results
in lifting the degeneracies. The perturbation-affected eigenvalues
have, therefore, real part approximately given by $\Ga c_k$.

\textit{Spectra of the stripes.}---The stripe associated to $c_0=0$ is
described by the equation for $R_0$ considered in
\cite{2018ZenoDynamics,ZenoStatics}.  We review this equation and evaluate the
corresponding spectrum in \cite{SM}.  The other Zeno stripes are
associated with the nonzero eigenvalues of the dissipator
(\ref{DissEig}): $c_1=c_2=-1/2$ and $c_3=-1$. In the following, we
consider the eigenvalue $c_3$. The analysis of the degenerate
eigenvalue $c_1=c_2$ is similar and detailed in \cite{SM}.

To evaluate the $O(1)$ terms of Eq.~(\ref{statement}) for $k=3$, we
need the operators $U_3$ and $W_3$ (\ref{DefUWnd}).  The only nonzero
coefficients $A_{n,3,3}$ and $B_{n,3,3}$ with $n>0$ are
$A_{3,3,3}=B_{3,3,3} = - \mu$ and we find
$U_3=W_3= g_0- \mu g_3^\dagger = \sum_{j=1}^{N-1} h_{j,j+1} - \mu (J
\vec{n}_0)\cdot \vec{\sigma}_1$, where
$h_{j,j+1}=\vec{\sigma}_j \cdot (J \vec{\sigma}_{j+1})$ is the local
density of the Hamiltonian $H$.  Comparing $U_3$ and $h_D$, we see
that they differ just by the sign of the local field acting on site
$1$. It can be shown that $h_D$ and $U_3$ are, therefore,
isospectral~\cite{SM}.  According to (\ref{EigenvaluesRk}), the
corresponding Liouvillian eigenvalues are
\begin{align}
  \la_{3,\al,\be} = -\Ga  +i (\eps_\al- \eps_\be  ) + O(1/\Ga)
  \label{ResLaStripe3}
\end{align}
and the corresponding eigenvectors are
$\psi_{3,\al, \be} = \ket{\psi_3} \otimes \ket{\al}\bra{\be} +
O(1/\Ga)$, where $\ket{\al}$ is an eigenvector of $U_3$ with
eigenvalue $\eps_\al$.  The corrections $O(1/\Ga)$ are evaluated
according to Eq.~(\ref{CorrectionEigenvaluesRk}) for $\be\neq\al$. The
case $\be=\al$ is similar to the calculation done for $c_0=0$ and is
detailed in \cite{SM}.

 \begin{figure}
   \includegraphics[width=\columnwidth]{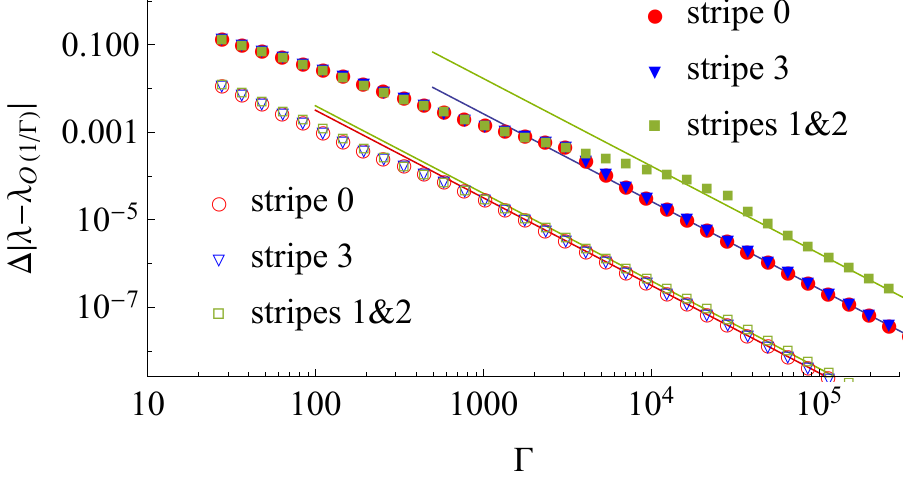} \centering
   \caption{Standard deviation of the modulus of the
     difference between numerically obtained Liouvillian eigenvalues
     and our perturbative prediction as a function of $\Ga$,
     separately for each stripe.  The set of data corresponding to
     empty symbols is obtained with parameters as in
     Fig.~\ref{Fig-LiuAllGamma05-20} (pure target state). The set with
     filled symbols corresponds to a mixed target state with
     parameters: $N=4, \vec{J}=(1,1.7,-0.137)$, $\varphi=0$,
     $\theta=2\pi/7$, $\mu=-0.7$.  The straight lines are
     $(\Ga_c/\Ga)^2$ with, from top to bottom,
     $\Ga_c=129,~51,~6.3,~5.6$. }
   \label{Error}
 \end{figure}

 Figure \ref{Error} shows, stripe by stripe, the standard deviation of
 the error obtained by comparing the numerically computed Liouvillian
 eigenvalues with our perturbative eigenvalues, order $O(1/\Ga)$
 included.  As expected, this error behaves like $(\Ga_c/\Ga)^2$ for
 $\Ga$ sufficiently large, with $\Ga_c$ possibly different for the
 various stripes depending on the parameters chosen.  The value of
 $\Ga_c$ can be used as an indicator of an onset of the Zeno regime,
 characterized by the appearance of stripes in
 Fig.~\ref{Fig-LiuAllGamma05-20}.  From Fig.~\ref{Error} we also see
 that the Zeno regime is reached easier for larger boundary gradient
 $|\mu|$.

\begin{figure}
  \includegraphics[width=\columnwidth]{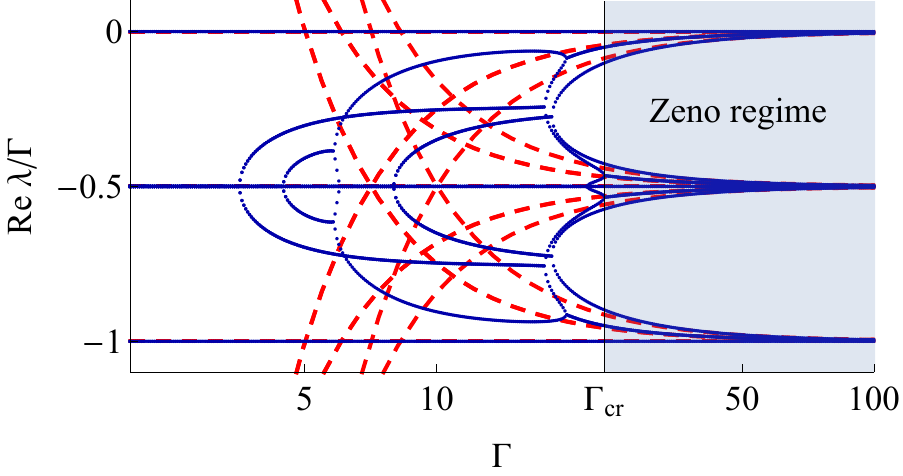} \centering
  \caption{Rescaled real part of all Liouvillian
    eigenvalues versus $\Ga$, for $N=1$.  Parameters:
    $\vec{J}=(1,2.3,-0.61)$, $\varphi=\theta=0$ and $\mu=1$. Dashed
    lines show the near Zeno-limit predictions detailed in
    \cite{SM}. The vertical line at $\Ga_\mathrm{cr}$ marks the
    location of the rightmost branching points where Zeno regime sets
    in.}
  \label{Fig-bifurcations}
\end{figure}

 Our Zeno-limit expansion for an eigenvalue $\la$ is applicable if the
 dissipation $\Ga$ is much larger than the inverse radius of
 convergence of the $1/\Ga^{k}$ perturbative series for $\la$. The
 global radius of convergence, valid for all Liouvillian eigenvalues,
 is problem specific.  In Fig.~\ref{Fig-bifurcations} we show, as a
 function of $\Gamma$, the real part of all Liouvillian eigenvalues of
 a Heisenberg chain with $2$ spins, the first spin being targeted by a
 $z$-polarizing dissipation.  Depending on the anisotropy, we find up
 to eight exceptional points, where two or more eigenvalues
 coalesce~\cite{EP-Kato, EP-Review,EP-Heiss,EP-Hatano}.  Fully
 analytical Zeno regime sets in beyond the rightmost branching point,
 see Fig.~\ref{Fig-bifurcations}.
 
Let us summarize our findings. The eigenvalues of a Liouvillian with a
locally acting dissipator at large dissipation strength $\Ga$ are
arranged in a set of stripes, see Fig.~\ref{Fig-LiuAllGamma05-20},
indicating the existence of a hierarchy of relaxation timescales in
the system~\cite{HyerarchyRelaxationTimescales}.  The number of
stripes coincides with the number of different eigenvalues of the
Lindblad dissipator $\mathcal{D}$ in (\ref{L}).

The width of the stripes scales as $1/\Ga$ and the distance between
the stripes scales as $\Ga$.  The vertical extension of the stripes
does not depend on $\Ga$ and is of the order of the norm $||H||$ of
the coherent part of the Liouvillian (\ref{L}). The position of the
stripes on the real axis is $\RE \la = c_k \Ga +O(1/\Ga)$ where $c_k$
are the eigenvalues of the dissipator (\ref{EqCalDeig}). Each stripe
corresponding to a nondegenerate $c_k$ contains $d_1^2$ Liouvillian
eigenvalues, where $d_1=\dim {\cal H}_1$ is the dimension of that part
of Hilbert space which is not affected directly by the dissipation.
Emergence of stripes can be viewed as a hallmark of a quantum Zeno
regime.

We derived linear spectral problems for the dissipation-projected
Liouvillian, for each relaxation mode $c_k$, and outlined a complete
solution of the eigenvalue problem via a perturbative analysis.  We
demonstrated our general results in the case of dissipation acting on
a single boundary qubit of an anisotropic Heisenberg spin chain. For
this case, we obtained explicit expressions for eigenvalues and
eigenvectors of the problem near the Zeno regime.  The solutions are
given in terms of spectral data of a dissipation-projected Hamiltonian
and other similar Hamiltonians, these being much simpler objects than
the original Liouvillian.  Our method is straightforwardly applicable
to the $XYZ$ model with dissipation acting on both boundaries, thus
creating boundary gradients \cite{2020ZenoPRL,2020ZenoPRE}, which play
a prominent role in studies of quantum transport
\cite{2020BertiniQuantumTransport}.  All the auxiliary Hamiltonians
have the form of an open $XYZ$ spin chain with boundary fields and are
integrable~\cite{OffDiagonal}.

To derive our results we used several assumptions: (i)
diagonalizability of the dissipator~(\ref{EqCalDeig}), (ii) uniqueness
of its kernel, (iii) absence of anomalous scaling of the gaps in the
spectrum of the Liouvillian, including the Liouvillian gap. A
generalization of our results is, in principle, straightforward.  We
expect the emergence of striped structure and scaling of the stripes
in the Zeno limit to be qualitatively correct also for degenerate
kernels, e.g., for those resulting from Hermitian Lindblad operators
\cite{RandomLiuLindbladOperatorsHermitian}.

Our explicit results shed a light on the intrinsic properties of an
isolated system coupled strongly to the environment, and make its
study almost analytically affordable.

\begin{acknowledgments}
  V.P. gratefully acknowledges financial support from the Deutsche
  Forschungsgemeinschaft through DFG project KL 645/20-1 and thanks
  the Department of Physics of Sapienza University of Rome for
  hospitality and financial support.
\end{acknowledgments}


%

\cleardoublepage
\newpage
\pagenumbering{arabic} \setcounter{page}{1} \onecolumngrid
\appendix

\begin{center}{\large\bf Supplemental Material for \\
Full spectrum of open dissipative quantum systems in the Zeno limit}
\end{center}

\begin{center}
  Vladislav Popkov and Carlo Presilla
\end {center}

\setcounter{equation}{0}
\renewcommand{\theequation}{S\arabic{equation}}
\setcounter{figure}{0}
\renewcommand{\thefigure}{S\arabic{figure}}

\section{Lamb shift Hamiltonian and effective dissipator
  of Eq.~(\ref{LMEeffective0})}

Assume that the kernel of $\mathcal{D}$ is one-dimensional, i.e., its
0 eigenvalue, $ \mathcal{D}[\psi_0] =0 $, is nondegenerate and
$\mathcal{D}$ is diagonalizable, i.e., there exists a basis
$\{\psi_k\}$ (not necessarily orthogonal) such that
$ \mathcal{D}[\psi_k] =c_k \psi_k $.  Let $\{\vfi_k\}$ be a
complementary basis, trace-orthonormal to the basis $\{\psi_k\}$,
$\tr(\vfi_k \psi_j)=\delta_{k,j}$. In Ref.~\cite{2018ZenoDynamics}
it has been shown that
\begin{align}
  &\tilde{H}_a = \sum_{m>0} \sum_{n>0} \beta_{m,n}
    g_m^\dagger g_n,
    \label{Dyson2-W.H}
  \\
  &\tilde{\mathcal{D}}[\cdot] = \sum_{m>0}\sum_{n>0}
    \gamma_{m,n} \left( g_n \cdot g_m^\dagger -
    \frac{1}{2} g_m^\dagger g_n \cdot -
    \frac{1}{2} \cdot g_m^\dagger g_n \right),
    \label{Dyson2-W.D}
\end{align}
where
$g_k = \trzero ((\psi_k \otimes I_{\mathcal{H}_1})
H) \label{Dyson2-hk}$ are the operators in Eq.~(\ref{gk}) and
$\gamma_{m,n}=Y_{m,n}+Y_{n,m}^*$ and $\beta_{m,n}=(Y_{m,n}-Y_{n,m}^*)/(2i)$ with
$Y_{m,n}=-\tr \left( \vfi_m^\dagger \vfi_n \psi_0 \right)/c_m^*$ are
the elements of two matrices which are, respectively, positive and
Hermitian. Note that the dissipation-projected Hamiltonian of
Eq.~(\ref{LMEeffective0}) is $h_D=g_0$.

For the dissipator
$\mathcal{D} = ((1+\mu)/2) \mathcal{D}_1 + ((1-\mu)/2) \mathcal{D}_2$
with $\mathcal{D}_1$ and $\mathcal{D}_2$ given by
Eq.~(\ref{DefLMEDissipator}), we have $\beta_{m,n}=0$ and
$\gamma_{m,n} = \gamma_m \delta_{m,n}$, where $\gamma_1=(1+\mu)/2$,
$\gamma_2=(1-\mu)/2$ and $\gamma_3=(1-\mu^2)/4$.

\section{Proof of Statement: nondegenerate eigenvalues}
We start introducing the spectral projection
 $P_k$ according to
 \begin{align}
   P_k X
   &= \psi_k \otimes X_k, \qquad X_k=
     \trzero \left( ( \vfi_k \otimes I_{\mathcal{H}_1})
     X  \right ). 
 \end{align}
 We have $P_k P_m = \de_{k,m} P_m$ and
 $ P_k \rho(\tau) = \psi_k \otimes R_k(\tau)$.  From Eq.~(\ref{L}),
 scaling the time by $\tau= t/\Ga$, we find
 $d\rho(t)/dt=\cL0[\rho(t)]+K[\rho(t)]$ with
 $\cL0[\cdot]=\mathcal{D}[\cdot]$ and
 $K[\cdot] = -(i/\Gamma) [H,\cdot]$.  If we now apply the Liouvillian
 propagator $\eps_t = \exp\mathcal{L} t$ on $P_k$ with $k>0$, we can
 use the Dyson expansion with respect to the small perturbation $K$
 and obtain
 \begin{align}
   \eps_t P_k
   =&\  e^{c_k t} \left( P_k + P_k K P_k t \right) + \frac{1}{c_k} P_0 K P_k
      (e^{c_k t} -1)
      \nonumber\\
    &+\!\!\!\!\!\!\!
      \sum_{m>0,~m\neq k}  \frac{  e^{c_m t}}{c_k-c_m} P_m K P_k
      (e^{(c_k - c_m) t} -1) + O(K^2). 
      \label{EqEPn }
 \end{align}
 The term $P_0 K P_k$ describes the flow towards the dissipation-free
 subspace; as expected, its norm is of order $1/\Ga$ due to presence
 of $K$. The term in (\ref{EqEPn }) containing $P_m K P_k$ describes
 the intra-sector flow
 $\psi_k \otimes R_k(0) \rightarrow \psi_m \otimes R_m(t) $, and is at
 most of order $1/\Ga$ at any time.  Finally, the inter-sector flow
 $\psi_k \otimes R_k(0) \rightarrow \psi_k \otimes R_k(t) $ is given
 by the first two terms, namely,
 \begin{align}
   &P_k \eps_t P_k =  e^{c_k t} \left( P_k + P_k K P_k t \right) + O(K^2).  
     \label{EqPnEPn}
 \end{align} 
 The evolution $R_k(0) \rightarrow R_k(t)$ resulting from
 Eq.~(\ref{EqPnEPn}) can be cast in differential form by using
 $dR_k(t)/dt = \lim_{t \rightarrow 0} (R_k(t)-R_k(0))/t$. Applying
 $P_k \eps_t P_k$ on $\rho(t)$ we find
 $\psi_k \otimes dR_k(t)/dt =c_k \psi_k \otimes R_k + P_k K P_k
 \rho(t)$.  Scaling back the time by $t=\Ga \tau$, after some algebra
 we get
 \begin{align}
   \frac{d R_k(\tau)}{d \tau}
   =&\ \Ga c_k R_k(\tau) + i [g_0, R_k(\tau)] 
   \nonumber\\
    &+
   i \sum_{n>0} 
   \left( \tr(\vfi_k \vfi_n^\dagger \psi_k) g_n^\dagger R_k(\tau) -  
   R_k(\tau)  \tr(\vfi_k \psi_n \vfi_k^\dagger)  g_n^\dagger\right)
      + O(1/\Ga),
   \label{LinearProblemRk}
 \end{align}
 which, by virtue of Eq.~(\ref{DefUWnd}), is Eq.~(\ref{statement}) up
 to terms $O(1)$.  The $O(1/\Ga)$ corrections can be obtained by
 accounting for the next, second order term of the Dyson expansion,
 see later.

 \section{Proof of Statement: degenerate eigenvalues}
 Suppose that there exists a degenerate dissipator eigenvalue with
 degeneracy $\mathrm{deg}$, let's say,
 $c_k=c_{k+1}= \dots = c_{k+\mathrm{deg}-1}$.  Equation~(\ref{EqEPn })
 is not applicable directly, since there would be a pole singularity
 in the terms $1/(c_k-c_m)$.  In order to eliminate this singularity,
 we group together the respective spectral projections $P_k$, defining
 ${\bf P}= P_k+P_{k+1} + \dots +P_{k+\mathrm{deg}-1}$.  One can check
 that Eq.~(\ref{EqEPn }) with the substitution
 $(P_k,P_{k+1},\dots,P_{k+\mathrm{deg}-1}) \rightarrow {\bf P}$
 remains valid provided the sum over $m$ has the constraint
 $m\neq k,k+1,\dots,k+\mathrm{deg}-1$, and we obtain
 ${\bf P} \eps_t {\bf P} = e^{c_k t} \left( {\bf P} + {\bf P} K {\bf
     P} t \right) + O(K^2)$. For the equation of motion of the
 components $R_k(\tau),R_{k+1}(\tau),\dots,R_{k+\mathrm{deg}-1}(\tau)$,
 we get
 \begin{align}
   \frac{ d R_k(\tau)}{d \tau}  
   =&\  \Ga c_k R_k(\tau) + i [g_0, R_k(\tau)] 
      \nonumber\\
    &+ i \sum_{n>0} \sum_{s:c_s=c_k} 
      \left( \tr(\vfi_k \vfi_n^\dagger \psi_s) g_n^\dagger R_s(\tau) -  
      R_s(\tau) \tr(\vfi_k \psi_n \vfi_s^\dagger) g_n^\dagger\right)
      + O(1/\Ga),
      \label{LinearProblemRkDeg}
 \end{align} 
 which, by virtue of Eq.~(\ref{DefUW}),
is Eq.~(\ref{statement}) up to terms $O(1)$.

 \section{Proof of statement: Dyson expansion at second order}
 To obtain the $O(1/\Ga)$ terms in the equation of motion for $R_k(t)$
 we need to include in the Dyson expansion the terms of order 2 in the
 perturbation $K$.  The $O(K^2)$ term for the evolution projected onto
 the $R_k$ subspace is given by the operator
 $P_k \exp({\cal L}t) = P_k \eps_t$. Recalling that
 $P_k \rho= \psi_k \otimes R_k$, we have
 \begin{align}
   &\psi_k \otimes R_k(t) = P_k \rho(t) = P_k \eps_t \rho(0)
     = \sum_{j} P_k \eps_t  P_j \rho(0).
 \end{align}
 In differential form we have $d R_k(\tau)/d\tau = \Ga d R_k(t)/dt $, i.e.,
 \begin{align}
   &\psi_k \otimes \frac {d R_k(\tau)}{d \tau} =\Ga \lim_{t\rightarrow 0}
     \frac{ \sum_{j} P_k \eps_t  P_j \rho(0) - P_k \rho(0)}{t} .
     \label{eq:eqMotionRk}
 \end{align}

 It turns out that the $O(1/\Ga)$ contribution to the equation of
 motion (\ref{eq:eqMotionRk}) for $R_k(t)$ are given only by the terms
 $P_k \eps_t P_s \rho(0)$, with $c_s=c_k$, while the terms
 $P_k \eps_t P_n \rho(0)$ with $c_n \neq c_k $ give no $O(1/\Ga)$
 contribution.  The Dyson expansion for $P_k \eps_t P_s$
 with $c_s=c_k$ yields
 \begin{align}
   &P_k \eps_t P_s = \de_{s,k} P_k + O(K)+
     t e^{c_k t} 
     \sum_{n:c_n\neq c_k}
     \frac{1}{c_k-c_n} P_k K P_n K P_s,
 \end{align}
 where the $O(K)$ terms are those calculated before. At the leading
 order in time, $ t e^{c_k t} = t+ O(t^2)$. In differential form,
 the respective terms for $R_k(\tau)$ are given by
 \begin{align}
   &\psi_k \otimes \frac{d R_k(\tau)}{d \tau} =
     O(1)+ \Ga
     \sum_{n:c_n\neq c_k}
     \sum_{s:c_s=c_k}
     \frac{1}{c_k-c_n} P_k K P_n K P_s \rho(\tau).
 \end{align}

 Using the following formulas
 \begin{align}
   &\rho= \sum_k \psi_k \otimes R_k,\\
   &\tr(\vfi_k \psi_n)= \de_{k,n},\\
   &P_k A= \psi_k \otimes \tr (A \vfi_k), \qquad
     P_k \rho= \psi_k \otimes R_k, \\
   &H=\sum_m \vfi_m \otimes  g_m = \sum_m \vfi_m^\dagger \otimes  g_m^\dagger,\\
   &K A = - \frac{i}{\Ga} [H,A],
 \end{align}
 we calculate the term $P_k K P_n K P_s \rho$, step by step, as follows
 (summation over repeated indices $m$ is implied)
 \begin{align*}
   P_n K P_s \rho &=   - \frac{i}{\Ga} P_n [H, \psi_s \otimes R_s] \\
   &= - \frac{i}{\Ga}P_n  (H ( \psi_s \otimes R_s ) -(\psi_s \otimes R_s) H)\\
   &= - \frac{i}{\Ga} P_n  \left(
     ( \vfi_m \otimes  g_m)  ( \psi_s \otimes R_s ) - (\psi_s \otimes R_s)
     (\vfi_m^\dagger \otimes  g_m^\dagger) \right)\\
   &= - \frac{i}{\Ga} P_n  \left(
     ( \vfi_m \psi_s \otimes  g_m R_s ) -
     (\psi_s \vfi_m^\dagger \otimes R_s g_m^\dagger) \right)\\
   &= - \frac{i}{\Ga} \psi_n  \otimes  \left(
     \tr ( \vfi_n \vfi_m \psi_s )   g_m R_s  -
     \tr (\vfi_n \psi_s \vfi_m^\dagger) R_s g_m^\dagger \right)\\
   &= - \frac{i}{\Ga} \psi_n  \otimes  \left(
     C_{m,s,n} g_m R_s  - A_{m,s,n} R_s g_m^\dagger \right),
 \end{align*}
 and then (now, summation over repeated indices $m$ and $z$ is implied)
 \begin{align*}
   &P_k K (P_n K P_s \rho)  \\
   &= - \frac{1}{\Ga^2} P_k \left(
     C_{m,s,n}  [H, \psi_n  \otimes     g_m R_s] - 
     A_{m,s,n} [ H,   \psi_n  \otimes   R_s g_m^\dagger)]
     \right)\\
   &= - \frac{1}{\Ga^2} P_k (
     C_{m,s,n}  [ \vfi_z^\dagger \otimes  g_z^\dagger) ,
     (\psi_n  \otimes     g_m R_s] 
     -A_{m,s,n} [ \vfi_z \otimes  g_z) ,
     (\psi_n  \otimes     R_s g_m^\dagger]
     )\\
   &= - \frac{1}{\Ga^2} P_k (
     C_{m,s,n}  ( \vfi_z^\dagger \psi_n \otimes  g_z^\dagger
     g_m R_s - \psi_n \vfi_z^\dagger  \otimes g_m R_s g_z^\dagger )
     -A_{m,s,n} (  \vfi_z \psi_n   \otimes
     g_z   R_s g_m^\dagger -   \psi_n  \vfi_z \otimes R_s g_m^\dagger g_z)
     \\
   &=- \frac{1}{\Ga^2} \psi_k   \otimes (
     C_{m,s,n} B_{z,n,k}  g_z^\dagger    g_m R_s  
     -C_{m,s,n} A_{z,n,k}   g_m R_s g_z^\dagger
     -A_{m,s,n}  C_{z,n,k} g_z   R_s g_m^\dagger 
     + A_{m,s,n} C_{k,n,z}   R_s g_m^\dagger g_z ) \\
   &=- \frac{1}{\Ga^2} \psi_k   \otimes 
     (
     -(C_{m,s,n} A_{z,n,k} +A_{z,s,n}  C_{m,n,k}  )  g_m R_s g_z^\dagger 
     + C_{m,s,n} B_{z,n,k}  g_z^\dagger    g_m R_s 
     + A_{z,s,n} C_{k,n,m}   R_s g_z^\dagger g_m
     ) .
 \end{align*}
 In passing from the second-last line to the last one, we exchanged the
 summation indices $m\leftrightarrow z$ in half of the terms.
 Finally, denoting
 \begin{align}
   \ga^{n,s,k}_{m,z} &= C_{m,s,n} A_{z,n,k} +A_{z,s,n}  C_{m,n,k},
   \\
   \eps^{n,s,k}_{z,m} &= C_{m,s,n} B_{z,n,k},
   \\
   \de^{n,s,k}_{z,m} &= A_{z,s,n} C_{k,n,m},
 \end{align}
 and multiplying by $\Ga$, we obtain the $O(1/\Ga)$ terms of
 Eq.~(\ref{statement}).

 \section{Equivalence of two open spin chains with flipped boundary
   fields }
 Suppose that we have two operators $ f_{\pm}$ of the form
 \begin{align}
   & f_{\pm} = \sum_{j=1}^{N-1} \sum_{\al=x,y,z} J_\al \sigma_j^\al
     \sigma_{j+1}^\al
     \pm \sum_{\al=x,y,z}  n_\al \sigma_1^\al,
 \end{align}
 where $J_\al,n_\al$ are some constants.  Let us choose a
 representation in which the boundary term becomes diagonal, by an
 appropriate rotation of the basis,
 $\sum_{\al=x,y,z} n_\al \sigma_1^\al = A \tsi_1^z$.  Under this
 transformation the operators $ f_{\pm}$ take the form
 \begin{align}
   f_{\pm} = \sum_{j=1}^{N-1} \sum_{\al,\be=x,y,z} K_{\al\be} \tsi_j^\al
   \tsi_{j+1}^\be   \pm A \tsi_1^z,  
 \end{align}
 where $K_{\al\be}$ and $A$ are constants.  Then, the unitary operator
 \begin{align}
   &U = \bigotimes_{j=1}^{N} \tsi_j^x, \quad U^2=I,
     \label{UnitaryTransformation}
 \end{align}
 transforms $f_{+}$ into $f_{-}$ and vice versa,
 \begin{align}
   & f_{\pm} = U f_{\mp} U,
 \end{align} 
 which follows from $\tsi_j^x \tsi_j^z \tsi_j^{x,y} = - \tsi_j^z $ and
 $\tsi_j^x \tsi_j^x \tsi_j^x =\tsi_j^x$.

 \section{The $XYZ$ spin chain: spectrum associated to the
  dissipator eigenvalue $c_0=0$.}
 This is the stripe closest to the origin in
 Fig.~\ref{Fig-LiuAllGamma05-20}.  The equation for $R_0$ was obtained
 in \cite{2018ZenoDynamics}. It has the Lindblad form
 (\ref{LMEeffective0}) with $\tilde{H}_a=0$,
 \begin{align}
   \label{hD}
   h_D = \sum_{j=1}^{N-1} \vec{\sigma}_j \cdot (\hat{J} \vec{\sigma}_{j+1})
   +\mu (\hat{J}\vec{n}_0) \cdot \vec{\sigma}_1,  
 \end{align}
 where $\hat{J}=\mathrm{diag}(J_x,J_y,J_z)$, and effective dissipator
 $ \tilde{\mathcal{D}} [R_0] = \sum_{p=1}^{3} ( {\tilde{L}_p R_0
   \tilde{L}_p ^\dagger} - {\frac{1}{2} \tilde{L}_p ^\dagger
   \tilde{L}_p R_0} - {\frac{1}{2}R_0 \tilde{L}_p ^\dagger
   \tilde{L}_p} )$ with
 \begin{align*}
   &\tilde{L}_1 =
     \sqrt{2(1+ \mu)} \left(  \hat{J}(\vec{n}_0'-i \vec{n}_0 )\right)
     \cdot \vec{\si}_1,
   \\
   &\tilde{L}_2 =
     \tilde{L}_1^\dagger \sqrt{(1- \mu)}/  \sqrt{(1+ \mu)},
   \\
   &\tilde{L}_3 =
     \sqrt{(1- \mu^2)/2} \left(  \hat{J}\vec{n}_0   \right) \cdot \vec{\si}_1,
 \end{align*}
 where
 $\vec{n}_0 = \vec{n}(\th,\vfi) \equiv (\sin \th \cos \vfi,\sin \th
 \sin \vfi,\cos \th)$, and
 $\vec{n}_0'= \vec{n}(\frac{\pi}{2}-\th,\vfi+ \pi)$,
 $\vec{n}_0= \vec{n}(\frac{\pi}{2},\vfi+ \frac{\pi}{2})$.

 Neglecting $O(1/\Ga)$ corrections,
 eigencomponents of the matrix $R_0$ have form
 $\ket{\psi_0} \otimes \ket{\al}\bra{\be}$, with respective
 eigenvalues $ \la_{0,\al,\be} = i (\eps_\be- \eps_\al ) + O(1/\Ga)$,
 where $h_D \ket{\al} = \eps_\al \ket{\al}$.  Note that the
 eigenvalues $\eps_\al$ are real because $h_D$ is Hermitian.
 Including the $O(1/\Ga)$ corrections, the eigenvalues
 $\la_{0,\al,\be}$ are given by the perturbative formula
 \begin{align}
   \la_{0,\al,\be} =&\ i (\eps_\be- \eps_\al  ) 
                      +\  \frac{1}{\Ga}
                      \sum_{p=1}^3 \bigg( 
                      \bra{\al} \tilde{L}_p \ket{\al}
                      \bra{\be} \tilde{L}_p^\dagger\ket{\be}
                      -\frac{1}{2} \bra{\al} \tilde{L}_p^\dagger
                      \tilde{L}_p \ket{\al} 
                      -\frac{1}{2} \bra{\be}  \tilde{L}_p^\dagger
                      \tilde{L}_p \ket{\be}
                      \bigg) .
                      \label{CorrectionEigenvaluesR0}
 \end{align}
 The above $O(1/\Ga)$ corrections are valid only for eigenvalues
 nondegenerate at the zeroth order, i.e., for $\al \neq \be$.  For
 degenerate eigenvalues $\la_{0,\al,\al}$, to resolve the degeneracy
 we write down equations for
 $\nu_\al(\tau) = \bra{\al} R_0(\tau) \ket{\al}$ using
 Eq.~(\ref{LMEeffective0}).  We obtain (see also
 \cite{ZenoStatics}) a classical Markov process
 $d \nu_\al(\tau)/d \tau = \Ga^{-1} \sum_{\be} M_{\al\be}\
 \nu_{\be}(\tau)$, where $M$ is the stochastic matrix with elements
 $M_{\al\be} = \sum_{p} | \bra{\al} \tilde{L}_p \ket{\be} |^2$, for
 $\al\neq\be$, and $M_{\al\al} = -\sum_{\be \neq \al} M_{\be \al}$.
 The eigenvalues of $M$, namely, $M \ket{P_\al}= \mu_\al \ket{P_\al}$,
 determine the $O(1/\Ga)$ corrections to the $2^{N-1}$ degenerate
 eigenvalues $\la_{0,\al,\al}$
 \begin{align}
   \la_{0,\al,\al} =  \frac{1}{\Ga} \mu_\al + O(1/\Ga^2) .
   \label{diagLa}
 \end{align}
 According to the Perron-Frobenius theorem, all eigenvalues
 $\mu_{\al}$ have a strictly negative real part, except for $\al=0$
 which is $\mu_0=0$.  This zero eigenvalue corresponds to an
 eigenvector $\ket{P_0}$ with real nonnegative entries $\nu_\al$. In
 the original quantum problem, the $\nu_\al$ have the meaning of
 eigenvalues of the reduced density matrix in the Zeno
 limit~\cite{2018ZenoDynamics}.  We remark that the $O(1/\Gamma)$
 corrections in Eqs.~(\ref{CorrectionEigenvaluesR0}) and
 (\ref{diagLa}) have strictly negative real part and, in addition, all
 $\mu_\al$ from Eq.~(\ref{diagLa}) are real, which is a highly
 nontrivial property.

\begin{figure}
   \includegraphics[width=1\columnwidth]{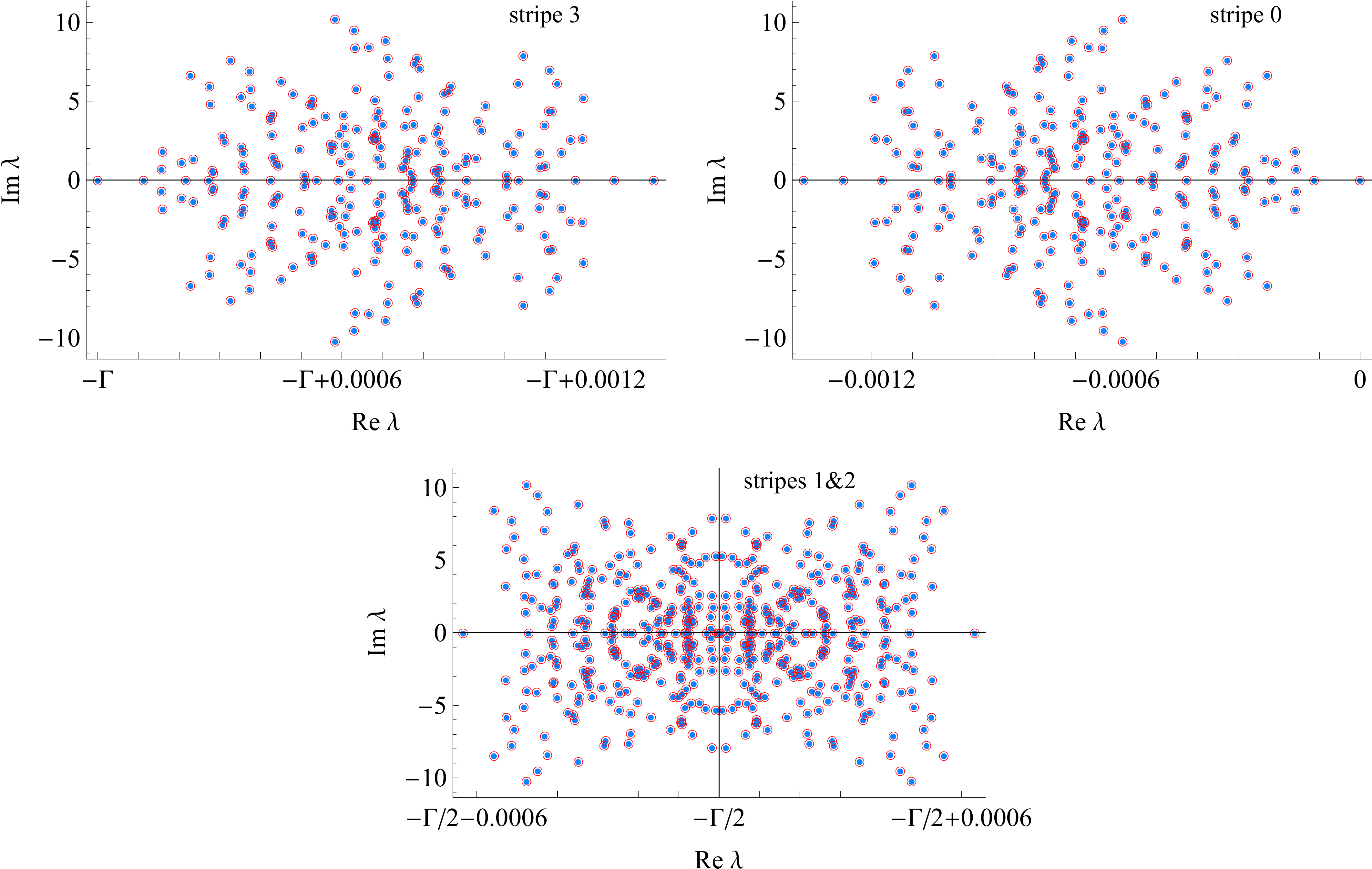}
   \caption{Complex eigenvalues of the Liouvillian
     belonging to the stripes 0, 3 and 1\&2 for $\Ga=8000$.
     Approximated eigenvalues (open red circles) are computed at order
     $1/\Gamma$ by Eqs.~(\ref{CorrectionEigenvaluesR0}) and
     (\ref{diagLa}) for stripe 0, Eqs.~(\ref{eq:la3Corrections_mu})
     and (\ref{Res3diag-corrections}) for stripe 3 and
     Eqs.~(\ref{l1ab}), (\ref{l2ab}) and (\ref{Res12diag-corrections})
     for stripes 1\&2, and compare very well with the exact numerical
     results (blue dots).  Parameters as in
     Fig.~\ref{Fig-LiuAllGamma05-20}.}
   \label{FigAllStripesComparison}
 \end{figure}

 In the top right panel of Fig.~\ref{FigAllStripesComparison},
 we compare the Liouvillian eigenvalues
 of this stripe evaluated numerically with those obtained by the above
 perturbative formulas. As expected according to Fig.~\ref{Error},
 for the chosen value $\Ga=8000$ we have an excellent agreement between the
 two sets of data.

 \section{The $XYZ$ spin chain: corrections $O(1/\Ga)$ for the
   spectrum associated to the nondegenerate dissipator eigenvalue
   $c_3=-1$}
 
 First of all, we note that for the $XYZ$ spin chain with dissipation at site 0
 the operators
 $g_k = \trzero ((\psi_k \otimes I_{\mathcal{H}_1})
H) \label{Dyson2-hk}$
are given by~\cite{2018ZenoDynamicsSM}, Eqs.~(41) and (42), 
\begin{align*}
  &g_1 =
    \left(  \hat{J}(\vec{n}_0'-i \vec{n}_0 )\right)
    \cdot \vec{\si}_1,
  \\
  &g_2 = g_1^\dag,
  \\
  &g_3 = \left(  \hat{J}\vec{n}_0   \right) \cdot \vec{\si}_1.
\end{align*}

The $O(1/\Ga)$ corrections $\de \la_{3,\al,\be}$ to the Liouvillian
 eigenvalues $\la_{3,\al,\be}= c_3 \Ga + i( \eps_\al - \eps_\be)$ are
 obtained from the second order of Dyson expansion and correspond to
 the terms $O(1/\Ga)$ of Eq.~(\ref{statement}).  By explicitly
 calculating the coefficients $\ga^{nsk}_{mz}$, $\eps^{nsk}_{zm}$ and
 $\de^{nsk}_{zm}$ with $s=k=3$ and $n =0,1,2$, we find
\begin{align}
  \frac{d {R_3}}{d \tau}
  &= \Ga c_3 R_3 + i (U_3 R_3 - R_3 W_3) + \frac{1}{\Ga}
    \left( (1+\mu) E_{2,1}[R_3] + (1-\mu) E_{1,2}[R_3] -
    \frac{1-\mu^2}{2}{\cal D}_{g_3}[R_3] \right),
    \label{EqR3-2}
    \end{align}
where
\begin{align}
  U_3
  &= V_3 = g_0 - \mu \ g_3^\dagger =
    \sum_{k=1}^{N-1} h_{k,k+1} - \mu (J \vec{n}_0)\cdot \vec{\sigma}_1,
    \label{eq:u3}
  \\
  E_{n,m}[X]
  &= g_n^\dagger g_n X + X  g_n^\dagger g_n  + 2 g_m X g_m^\dagger,
  \\
  {\cal D}_g[X]
  &= g X g^\dagger - \frac{1}{2} g^\dagger   g X - \frac{1}{2} X g^\dagger g.
\end{align} 
For $\Ga$ large, the last term in Eq.~(\ref{EqR3-2}) can be treated as
a perturbation $V_3[R_3]$ of order $1/\Ga$.  The $O(1/\Ga)$
corrections to the Liouvillian eigenvalues are then obtained via the
standard perturbative formula
$\de \la_{3,\al,\be}= \bra{\al \be} \hat{V}_3 \ket{\al \be}$, where
$\hat{V}_3$ is the vectorized superoperator acting on the vectorized
reduced density matrix
$\ket{R_3}=\ket{\al,\be}=\ket{\al} \otimes \ket{\be}^*$ defined by
$\hat{V}_3\ket{R_3}=V_3[R_3]$.  We recall that $\ket{\al}$ and
$\ket{\be}$ are the eigenvectors of $V_3=V_3$,
\begin{align}
  U_3 \ket{\al} = \eps_\al \ket{\al}.
\end{align}
Note that $U_3$ is Hermitian and its eigenvalues $\eps_\al$ are
real.

To explicitly illustrate the evaluation of $\de \la_{3,\al,\be}$,
let's start considering the simplest case $\mu=1$.  By making the
substitution $R_3(\tau)= e^{c_3 \Ga \tau}r_3(\tau)$, we obtain
\begin{align}
  \frac{d {r_3}}{d \tau}
  &=  i(U_3 r_3 - r_3 U_3 ) + \frac{2}{\Ga} \left(
    g_2^\dagger g_2 r_3 + r_3  g_2^\dagger g_2
    + 2 g_1 r_3 g_1^\dagger \right)
    \nonumber\\
  &= {\cal L}_3^{(0)} [r_3] + V_3[r_3].
  \label{EqFor-r3}
\end{align} 
In the Zeno limit $\Ga \rightarrow \infty$, Eq.~(\ref{EqFor-r3}) for
$r_3$ is linearized in terms of modes $\ket{\al} \bra{\be}$.  In fact,
$U_3$ can be obtained from $h_D$ by flipping the
boundary term, therefore $h_D$ and $U_3$ are equivalent and have the
same set of eigenvalues $\eps_\al$.  It follows that, in an equivalent
representation, the solution of the eigenvalue problem for the
Liouvillian $\mathcal{L}^{(0)}_3[\cdot]$, namely,
$\mathcal{L}^{(0)}_3 [\psi _j] = \Lambda_j \psi_j$, is given by
$\psi_j = \ket{\al} \bra{\be}$ and
$\Lambda_j=i( \eps_\al - \eps_\be)$.

The expectation of an arbitrary superoperator of the form
$V[r_3] = Q r_3 W$ on the state
$\psi_j = \ket{\al}\bra{\be}$ can be calculated
in a vectorized form as
\begin{align}
  \bra{\psi_j} \hat{V} \ket{\psi_j}=
  \bra{\al}\otimes\bra{\be}^* ( Q \otimes W^t) \ket{\al}\otimes \ket{\be}^*
  = \bra{\al}Q\ket{\al}  \bra{\be}^* W^t \ket{\be}^*
  = \bra{\al}Q\ket{\al}  \bra{\be} W \ket{\be}.
\label{eq:Vjj-reduction}
\end{align} 
It follows that, accounting for
the corrections $O(1/\Ga)$, for $\mu=1$ we obtain
\begin{align}
  \la_{3,\al,\be}
  &= c_3 \Ga + i( \eps_\al - \eps_\be) +
    \bra{\psi_j} \hat{V}_3 \ket{\psi_j}
    \nonumber \\
  &=  -\Ga + i( \eps_\al - \eps_\be) +  \frac{2}{\Ga} 
    \left( \bra{\al} g_2^\dagger g_2 \ket{\al}   +
    \bra{\be}   g_2^\dagger g_2  \ket{\be} + 2  \bra{\al}  g_1 \ket{\al}  
    \bra{\be}  g_1^\dagger\ket{\be} \right).
\label{eq:la3Corrections}
\end{align} 
This result is immediately generalised to arbitrary $\mu$
\begin{align}
  \la_{3,\al,\be}
  =&\ -\Ga + i( \eps_\al - \eps_\be)
  \nonumber \\
  &+  \frac{1}{\Ga}  \left( (1+\mu) \left(
    \bra{\al} g_2^\dagger g_2 \ket{\al}   +
    \bra{\be}   g_2^\dagger g_2  \ket{\be} + 2  \bra{\al}  g_1 \ket{\al}  
    \bra{\be}  g_1^\dagger\ket{\be} \right)  \right.
    \nonumber \\
   &\qquad+ (1-\mu)  \left(
     \bra{\al} g_1^\dagger g_1 \ket{\al} +
     \bra{\be}   g_1^\dagger g_1  \ket{\be} + 2  \bra{\al}  g_2 \ket{\al}  
     \bra{\be}  g_2^\dagger\ket{\be} \right)
     \nonumber\\ 
   &\qquad+ \frac{1-\mu^2}{4}\left.  \left(
     \bra{\al} g_3^\dagger g_3 \ket{\al}   +
     \bra{\be}   g_3^\dagger g_3  \ket{\be} -  2 \bra{\al}  g_3 \ket{\al}  
     \bra{\be}  g_3^\dagger\ket{\be} \right) \right).
     \label{eq:la3Corrections_mu}
\end{align} 

The above perturbative formula can be applied only if the unperturbed
eigenvalue is nondegenerate. For $O(1)$ degenerate eigenvalues,
$\la_{3,\al,\al}= - \Ga $, the $O(1/\Ga)$ corrections must be found
in a different way.  In the Zeno limit, we have a stationary solution
$r_3( \infty) =\sum_\al \nu_\al \ket{\al} \bra{\al}$.  Taking
into account the $O(1/\Ga)$ terms, we can assume the finite-time $r_3(\tau)$
to have the same form but with coefficients $\nu_\al $ which depend on time,
$r_3( \tau) =\sum_\al \nu_\al(\tau) \ket{\al} \bra{\al}$.
Then, from Eq.~(\ref{EqFor-r3})
we have, for $\mu=1$,
\begin{align}
  \frac{d {\nu_\al}}{d \tau}
  = \frac{2}{\Ga} \sum_{\be} T_{\al,\be} \nu_\be, 
\end{align} 
where
\begin{align}
  T_{\al,\be}  = 2  w_{1,\al,\be},  \quad \be \neq \al,
                 \qquad
  T_{\al,\al}  = 2 \sum_\be w_{2,\be,\al} + 2 w_{1,\al,\al},
\end{align}
with 
\begin{align}
  w_{n,\al,\be} &= \left|  \bra{\al}  g_n \ket{\be} \right|^2.
\end{align}
For arbitrary values of $\mu$ we have, instead,
\begin{align}
  T_{\al,\be} = w_{\al,\be}(\mu),  \quad \be \neq \al,
                \qquad
  T_{\al,\al} = \sum_\be f_{\be,\al}(\mu) + w_{\al,\al}(\mu),
\end{align} 
with
\begin{align}
  w_{\al,\be}(\mu) &= (1+\mu)  w_{1,\al,\be} +
    (1-\mu) w_{2,\al,\be} - \frac{1-\mu^2}{4}  w_{3,\al,\be},\\
  f_{\be,\al}(\mu) &= (1+\mu) w_{2,\be,\al}  +
    (1-\mu) w_{1,\be,\al} + \frac{1-\mu^2}{4} w_{3,\be,\al}.
\end{align} 
By finding the eigenvalues $\mu_\al$ of the matrix $T$, we resolve the
degeneracy problem. In fact, in terms of the corresponding eigenvectors
$\tilde{\nu}_\al$ of $T$, we have
\begin{align}
  \frac{d {\tilde{\nu}_\al}}{d \tau}=
  \frac{2}{\Ga} \mu_\al \tilde{\nu}_\al,
\end{align} 
the set of the values $(2/\Ga) \mu_\al$ being the $1/\Ga$ correction
to the set of the degenerate eigenvalues $\la_{3,\al,\al}$,
\begin{align}
  \la_{3,\al,\al} = -\Ga + \frac{2}{\Ga} \mu_\al ,
  \qquad \al=1,2,\dots, 2^{N}. 
  \label{Res3diag-corrections}  
\end{align} 
Numerically, for the integrable $XYZ$ model, we find the matrix $T$ to
be equivalent to a symmetric real matrix, so that all its eigenvalues
$\mu_\al$ are real.  Since $c_3=-1$ is real too, the eigenvalues
(\ref{Res3diag-corrections}) lie on the real axis.

\section{The $XYZ$ spin chain: spectrum associated to the degenerate
  dissipator eigenvalue $c_1=c_2=-1/2$}

Equation~(\ref{statement}) for $k=1,2$ has the form
\begin{align}
  \frac{d R_k}{d\tau}
  =&\ \Ga c_1  R_k + i\sum_{s=1}^{2}
     \left( U_{k,s} R_s - R_s W_{k,s}\right)
     \nonumber\\
   &+\frac{1}{\Ga} \sum_{z>0} \sum_{m>0} \sum_{n=0,3}
     \sum_{s=1}^{2} \frac{1}{c_n-c_1}
     \left(
     -\ga^{n,s,k}_{m,z} g_m R_s g_z^\dagger
     + \eps^{n,s,k}_{z,m} g_z^\dagger g_m R_s
     + \de^{n,s,k}_{z,m} R_s g_z^\dagger g_m \right)
     \nonumber\\
  =&\ \Ga c_1  R_k + i\sum_{s=1}^{2}
     \left( U_{k,s} R_s - R_s W_{k,s}\right)
     \nonumber\\
   &+\frac{2}{\Ga} \sum_{z>0} \sum_{m>0} \sum_{s=1}^{2}
     \left(
     -\tilde{\ga}^{s,k}_{m,z} g_m R_s g_z^\dagger
     +\tilde{\eps}^{s,k}_{z,m} g_z^\dagger g_m R_s
     +\tilde{\de}^{s,k}_{z,m} R_s g_z^\dagger g_m 
     \right),
\end{align}
where
\begin{align}
  \tilde{\ga}^{s,k}_{m,z} &= \ga^{0,s,k}_{m,z}-\ga^{3,s,k}_{m,z},
  \\
  \tilde{\eps}^{s,k}_{m,z} &= \eps^{0,s,k}_{m,z}-\eps^{3,s,k}_{m,z},
  \\
  \tilde{\de}^{s,k}_{m,z} &= \de^{0,s,k}_{m,z}-\de^{3,s,k}_{m,z},
\end{align} 
with
\begin{align}
  \ga^{n,s,k}_{m,z} &= C_{m,s,n} A_{z,n,k} +A_{z,s,n}  C_{m,n,k},
  \\
  \eps^{n,s,k}_{z,m} &= C_{m,s,n} B_{z,n,k},
  \\
  \de^{n,s,k}_{z,m} &= A_{z,s,n} C_{k,n,m}. 
\end{align}
The only nonzero coefficients $\tilde{\ga}^{s,k}_{m,z}$,
$\tilde{\eps}^{s,k}_{m,z}$ and $\tilde{\de}^{s,k}_{m,z}$ are
\begin{align*}
  \tga^{1,1}_{1,1} &= \tga^{2,2}_{1,1}=1+\mu,
  \\
  \tga^{1,1}_{2,2} &= \tga^{2,2}_{2,2}=1-\mu,
  \\
  \teps^{2,2}_{2,2} &= -\teps^{1,1}_{1,1}= \mu,
  \\
  \teps^{2,1}_{1,2} &= 1-\mu, \qquad \teps^{1,2}_{2,1}=1+\mu,
  \\
  \tde^{1,1}_{2,2} &= -\teps^{2,2}_{1,1}= \mu,
  \\
  \tde^{2,1}_{1,2} &= 1+\mu, \qquad \tde^{1,2}_{2,1}=1-\mu.
\end{align*}

After the substitution $R_{1}(\tau)= e^{c_1 \Ga \tau}r_1(\tau)$ and
$R_{2}(\tau)= e^{c_1 \Ga \tau}r_2(\tau)$, we obtain the following
equations of motion for $r_1(\tau)$ and $r_2(\tau)$.
\begin{align}
  \frac{d {r_1}}{d \tau}
  =&\  i   \left( f_+ r_1 -r_1 f_- \right)
     \nonumber\\
   &+ \frac{2}{\Ga} \left(
     - (1+\mu) g_1 r_1 g_1^\dagger - (1-\mu) g_2 r_1 g_2^\dagger
     - \mu g_1^\dagger g_1 r_1  +\mu r_1 g_2^\dagger g_2 +\right.
     \nonumber\\
   &\qquad\left. +(1-\mu)  g_1^\dagger g_2 r_2
     + (1+\mu) r_2 g_1^\dagger g_2  \right),
     \label{EqFor-r1}
  \\
  \frac{d {r_2}}{d \tau}
  =&\  i   \left( f_- r_2 -r_2 f_+ \right)
     \nonumber\\
   &+ \frac{2}{\Ga} \left(-(1+\mu) g_1 r_2 g_1^\dagger
     - (1-\mu) g_2 r_2 g_2^\dagger +\mu g_2^\dagger g_2 r_2
     -\mu r_2 g_1^\dagger g_1 +\right.
     \nonumber\\
   &\qquad\left. +(1+\mu)  g_2^\dagger g_1 r_1
     + (1-\mu) r_1 g_2^\dagger g_1  \right),
     \label{EqFor-r2} 
\end{align} 
where, we recall that $g_0=h_D$,
\begin{align}
  f_\pm = g_0 \pm \frac{1\mp\mu}{2} g_3^\dagger
  =\sum_{j=1}^{N-1} h_{j,j+1} \pm (J \vec{n}_0)\cdot \vec{\sigma}_1.
\end{align}

At zeroth order in $1/\Gamma$, the eigenmodes of Eqs.~(\ref{EqFor-r1})
and (\ref{EqFor-r2}) are, respectively, $\ket{\al} \bra{\tilde{\be}}$
and $\ket{\tilde\al}\bra{\be}$, where $\ket{\al}$ and
$\ket{\tilde\al}$ are the eigenvectors of $f_+$ and $f_-$, namely,
$f_+\ket{\al}= \eps_{\al} \ket{\al}$ and
$f_-\ket{\tilde\al}= \eps_{\al} \ket{\tilde\al}$. Note that $f_+$ and
$f_-$, being related by a unitary transformation, have the same
eigenvalues. It follows that the zeroth order eigenvalues of the
Liouvillian are twice degenerate,
\begin{align}
  &\la_{1,\al,\be} = -\frac{\Ga}{2}  +i (\eps_{\al}- \eps_{\be}  )
    + O(1/\Ga),
    \label{ResLaStripe12}
  \\
  &\la_{2,\al,\be} = \la_{1,\al,\be}  + O(1/\Ga),
\end{align}
the respective eigenvectors being
$r_1^{(0)} =\ket{\al} \bra{\tilde\be}$ and
$r_2^{(0)}=\ket{\tilde\al}\bra{\be}$.  Note that the zeroth-order
eigenvalues $\la_{1,\al,\be}$ and $\la_{2,\al,\be}$ have a double
degeneracy for $\al \neq \be$ and a degeneracy $2^{N+1}$ for
$\al=\be$.

To obtain the $O(1/\Ga)$ corrections to the degenerate eigenvalues
$\la_{1,\al,\be}= \la_{2,\al,\be} = - \Ga/2 + i (\eps_\al - \eps_\be)
\equiv \Lambda_{\al \be}$, we substitute the Ansatz
$r_1 (\tau) = x_1(\tau) \ket{\al} \bra{\tilde\be}$ and
$r_2(\tau) = x_2(\tau) \ket{\tilde\al}\bra{\be}$ into
Eqs.~(\ref{EqFor-r1}) and (\ref{EqFor-r2}), obtaining the following
equations for $x_1(\tau)$ and $x_2(\tau)$
\begin{align*}
  \frac{d x_1}{d \tau}
  =&\ \Lambda_{\al \be} x_1 +  \frac{2}{\Ga} (V_{11} x_1  +  V_{12} x_2),
  \\
  \frac{d x_2}{d \tau}
  =&\ \Lambda_{\al \be} x_2+   \frac{2}{\Ga} (V_{21} x_1  +  V_{22} x_2),
\end{align*}
where
\begin{align*}
  V_{11}
  &= 
    -(1+\mu) \bra{\al} g_1 \ket{\al}
    \bra{\tilde\be} g_1^\dagger \ket{\tilde\be}  
    -(1-\mu) \bra{\al} g_2 \ket{\al}
    \bra{\tilde\be} g_2^\dagger \ket{\tilde\be}  
    -\mu \bra{\al} g_1^\dagger g_1 \ket{\al}
    +\mu \bra{\tilde\be} g_2^\dagger g_2 \ket{\tilde\be}, 
  \\
  V_{22}
  &=
    -(1+\mu) \bra{\tilde\al} g_1 \ket{\tilde\al}
    \bra{\be} g_1^\dagger \ket{\be}  
    -(1-\mu) \bra{\tilde\al} g_2 \ket{\tilde\al}
    \bra{\be} g_2^\dagger \ket{\be}  
    +\mu \bra{\tilde\al} g_2^\dagger g_2 \ket{\tilde\al}
    -\mu \bra{\be} g_1^\dagger g_1 \ket{\be},
  \\
  V_{12}
  &= 
    (1+\mu) \braket{\al}{\tilde\al}
    \bra{\be} g_1^\dagger g_2 \ket{\tilde\be} 
    +(1-\mu) \bra{\al} g_1^\dagger g_2 \ket{\tilde\al}
    \braket{\be}{\tilde\be},  
  \\
  V_{21}
  &= 
    (1+\mu) \bra{\tilde\al} g_2^\dagger g_1 \ket{\al}
    \braket{\tilde\be}{\be}
    +(1-\mu) \braket{\tilde\al}{\al}
    \bra{\tilde\be} g_2^\dagger g_1 \ket{\be}. 
\end{align*}  
The eigenvalues $v_{1},v_{2}$ of the matrix $V$ with elements $V_{ij}$
give the corrections to the eigenvalues
$ - \Ga/2 + i (\eps_\al - \eps_\be)$,
\begin{align}
  \la_{1\al \be} &=  - \Ga/2 + i (\eps_\al - \eps_\be) + \frac{2}{\Ga} v_1,
                   \label{l1ab}
  \\
  \la_{2\al \be} &=  - \Ga/2 + i (\eps_\al - \eps_\be) + \frac{2}{\Ga} v_2.
                   \label{l2ab}
\end{align}  


For degenerate eigenvalues
$\la_{1,\al,\al}= \la_{2,\al,\al} = -\Ga/2 $, the $O(1/\Ga)$
corrections have to be calculated in the following way.  In the Zeno
limit, the stationary solutions of Eqs.~(\ref{EqFor-r1}) and
(\ref{EqFor-r2}) are, respectively,
$r_1(\infty) = \sum_\al \nu_\al \ket{\al} \bra{\tilde\al}$ and
$r_2(\infty) = \sum_\al \mu_\al \ket{\tilde\al} \bra{\al}$.
Therefore, for $r_1(\tau)$ and $r_2(\tau)$ we may assume the form
$r_1(\tau) = \sum_\al \nu_\al(\tau) \ket{\al}\bra{\tilde\al}$ and
$r_2(\tau) = \sum_\al \mu_\al(\tau) \ket{\tilde\al} \bra{\al}$ with
coefficients $\nu_\al$ and $\mu_\al $ depending on time.  Inserting
these expressions into Eqs.~(\ref{EqFor-r1}) and (\ref{EqFor-r2}) and
writing down the equations for the components
$\bra{\al} r_1(\tau) \ket{\tilde\al} = \nu_\al(\tau)$ and
$\bra{\tilde\al} r_2(\tau) \ket{\al} = \mu_\al(\tau)$, we have
\begin{align}
  \frac{d {\nu_\al}}{d \tau}
  &= \frac{2}{\Ga}  \sum_{\be} 
    \left( T^{11}_{\al \be} \nu_\be +  T^{12}_{\al \be} \mu_\be \right),
  \\
  \frac{\partial {\mu_\al}}{\partial \tau}
  &= \frac{2}{\Ga}  \sum_{\be} 
    \left(  T^{21}_{\al \be} \nu_\be + T^{22}_{\al \be} \mu_\be \right),
\end{align}
where
\begin{align}
  T^{11}_{\al \be} &= w_1(\al,\be), \qquad \be \neq \al,\\
  T^{22}_{\al \be} &= w_2(\al,\be), \qquad \be \neq \al,\\
  T^{12}_{\al \be} &= w_{12}(\al,\be), \\
  T^{21}_{\al \be} &= w_{21}(\al,\be), \\
  T^{11}_{\al \al} &= w_1(\al,\al) + \sum_\be f(\al,\be), \\  
  T^{22}_{\al \al} &= w_2(\al,\al) + \sum_\be f(\al,\be),   
\end{align}
and
\begin{align}
  w_1(\al,\be)
  &= - (1+\mu) \bra{\al} g_1 \ket{\be}
    \bra{\tilde\be} g_1^\dagger \ket{\tilde\al}
    - (1-\mu) \bra{\al} g_2 \ket{\be}
    \bra{\tilde\be} g_2^\dagger \ket{\tilde\al},
  \\
  w_2(\al,\be)
  &= -(1+\mu) \bra{\tilde\al} g_1 \ket{\tilde\be}
    \bra{\be} g_1^\dagger \ket{\al}
    -(1-\mu) \bra{\tilde\al} g_2 \ket{\tilde\be}
    \bra{\be} g_2^\dagger \ket{\al},   
  \\
  f(\al,\be)
  &= \mu |\bra{\tilde\be} g_2 \ket{\tilde\al}|^2
    -\mu |\bra{\be} g_1 \ket{\al}|^2, 
  \\
  w_{12}(\al,\be)
  &= (1-\mu) \braket{\al}{\tilde\be}
    \bra{\be} g_1^\dagger g_2 \ket{\tilde\al}
    +(1+\mu) \braket{\be}{\tilde\al}
    \bra{\al} g_1^\dagger g_2 \ket{\tilde\be}, 
  \\
  w_{21}(\al,\be)
  &= (1-\mu) \braket{\tilde\al}{\be}
    \bra{\tilde\be} g_2^\dagger g_1 \ket{\al}
    +(1+\mu) \braket{\tilde\be}{\al}
    \bra{\tilde\al} g_2^\dagger g_1 \ket{\be}.  
\end{align}
By finding the eigenvalues $q_\al$ of the block matrix
\begin{align}
  &T=\left(
    \begin{array}{cc}
      T^{11}& T^{12}\\
      T^{21}& T^{22}
    \end{array}
              \right),
\end{align}
we resolve the degeneracy problem. The real eigenvalues with
$O(1/\Gamma)$ corrections, belonging to the degenerate eigenvalue
$c_1=c_2$ of the dissipator, are given by
\begin{align}
  \la_{1 \& 2,\al, \al}=  \frac{\Ga}{2} + \frac{2}{\Ga}  q_\al ,
  \qquad \al=1,2,\dots, 2 \times 2^{N}. 
  \label{Res12diag-corrections}
\end{align}
Numerically, we find that all the coefficients of the matrix $T$, as
the operators $f_\pm$, are $\mu$ independent and, therefore, the
corrections $q_\alpha$ in Eq.~(\ref{Res12diag-corrections}) are $\mu$
independent. This property is exceptional and probably connected with
the integrability of the $XYZ$ model.

\section{Properties of the auxiliary Markov Matrix $M_{ab}$}
It is well known that the eigenvalues of a generic stochastic matrix
are complex. Nevertheless, for our case example -- the XYZ model
with Zeno boundary dissipation -- all the eigenvalues happen to be real.

Here we prove this exceptional property, namely, that the eigenvalues
$\mu_a$ of the Markov matrix $M_{ab}$ in Eq (\ref{diagLa}) are all
real, for pure state boundary driving $\mu=1$.  We observe
(numerically) that the elements $M_{ab}$ of the Matrix Markov process,
\begin{align}
  \frac{d \nu_\al(\tau)}{d \tau} = \frac{1}{\Ga}
  \sum_{\be} M_{\al\be}\ \nu_{\be}(\tau),
  \label{MarkovProcess}
\end{align}
satisfy the so-called Kolmogorov condition
\begin{align}
&M_{ab} M_{bc} M_{ca} = M_{ac} M_{cb} M_{ba}, 
\label{Kolmogorov}
\end{align} 
with $a,b,c$ arbitrary and all different, if the targeted state at the
boundary is pure, i.e., for $\mu=1$. The Kolmogorov condition and
the positivity of the non-diagonal elements $M_{ab}$ entail
\begin{align*}
  &M_{ab}= s(a,b) \pi_b,
  \\
  &s(a,b) = s(b,a),
\end{align*} 
with $s(a,b)$ and $\pi_b$ real and positive. Introducing the diagonal
matrix $\hat{\pi}$ with elements $\pi_a$, we can write the Markov
matrix $M$ as
\begin{align*}
  M = \hat{\pi} S,
\end{align*} 
where $S$ is the matrix with non-diagonal elements $S_{ab}=s(a,b)$ and
$S_{aa} = M_{aa}/{\pi_a} $.  The above relation can be rewritten as
\begin{align*}
  \hat{\pi}^{-1/2} M  \hat{\pi}^{1/2}= \hat{\pi}^{1/2} S  \hat{\pi}^{1/2}.
\end{align*} 
Obviously, the RHS of the above equation is a real symmetric matrix,
since $S$ is a real symmetric matrix. Consequently,
$\hat{\pi}^{-1/2} M \hat{\pi}^{1/2}$ is also a real symmetric matrix,
i.e., the Markov matrix $M$ is equivalent to a real symmetric
matrix. Therefore, the eigenvalues $\mu_a$ of $M$ are all real.  It
follows that the $2^{N}$ Liouvillian eigenvalues belonging to the
first stripe~(\ref{diagLa}) lie, in the Zeno limit, on the real axis.

The same argument can be repeated for all stripes, and consequently,
all the Liouvillian eigenvalues of type $\la_{k,\al,\al}$ are, near
the Zeno limit, real. In total, for our $XYZ$ spin chain, there are
$4 \times 2^{N} = 2^{N+2}$ real Liouvillian eigenvalues, while all
the remaining Liouvillian eigenvalues $\la_{k,\al,\be}$, with
$\al \neq \be$ generically, i.e., in the absence of extra
degeneracies, have a nonzero imaginary part.

Finally, for $\mu \neq 1$ we observe numerically the same situation,
i.e., the eigenvalues of the Markov matrix $M$ (and its analogs for
the other stripes) are all real, so that the Zeno-limit Liouvillian
spectrum contains $2^{N+2}$ real entries.  Clearly, also in this case
$M$ must be equivalent to a Hermitian matrix. However, this fact can
no longer be explained by the Kolmogorov property (\ref{Kolmogorov}),
(equivalent to a detailed balance condition for the Markov rates
$w_{ab}=M_{ba}$) since this property is violated for $\mu \neq \pm 1$,
and the detailed balance condition $\pi_a w_{ab}= \pi_b w_{ba}$ is
consequently not satisfied. Further studies are required to clarify
this subtle issue.

  \section{Zeno limit for a problem with two qubits}

  Consider a problem (\ref{L}) with
  $H= \vec{\sigma}_0 \cdot (\hat{J} \vec{\sigma}_{1})$, where
  $\hat{J}=\mathrm{diag}(J_x,J_y,J_z)\equiv \mathrm{diag}(1,\ga,\De)$,
  and
  \begin{align}
    \mathcal{D}[\rho] =
    \si_0^+  \rho  \si_0^{-} - \frac{1}{2}    \si_0^{-} \si_0^{+} 
    \rho - \frac{1}{2}\rho  \si_0^{-} \si_0^{+} .
  \end{align}
  According to our general theory, the stripe closest to the imaginary
  axis, in the Zeno limit contains $4$ eigenvalues.  They are governed
  by the effective Hamiltonian (\ref{hD})
  \begin{align}
    h_D=\De \ \si^z
  \end{align}
  and by the effective Lindblad operator
  \begin{align}
    \tilde{L}_1 =
    -\left(
    \begin{array}{cc}
      0& 1+\ga\\
      1-\ga&  0
    \end{array}
             \right) .
  \end{align}
  The near-Zeno limit eigenvalues for the first stripe are given by
  Eq.~(\ref{CorrectionEigenvaluesR0}),
  \begin{align*}
    &\la_{0,1,1}=0,\\
    & \la_{0,1,2}= - 4\frac { 1+\ga^2} {\Ga} -2 \De i,\\
    &\la_{0,2,1}=  \la_{0,1,2}^*,\\
    &  \la_{0,2,2}= -8 \frac { 1+\ga^2} {\Ga}.
  \end{align*}
  Analogously, we obtain the other Liouvillian eigenvalues. The full
  set of $16$ Liouvillian eigenvalues $\la$ up to order $1/\Ga$ is
  given by
  \begin{align}
    & \la_{0,\al,\be}= \left\{ 0,  - 2\frac { \ga_{+}} {\Ga},
      - \frac { \ga_{+}} {\Ga}  \pm 2 \De i  \right\},
      \nonumber \\
    &\la_{1 \& 2,\al,\be}= \left\{ - \frac{\Ga}{2},
      - \frac{\Ga}{2}, - \frac{\Ga}{2}  \pm  \frac{2 \ga_{-}}{\Ga},
      - \frac{\Ga}{2} \pm \frac{8 \ga}{\Ga}   \pm 2  \De  i \right\},
      \label{la12-twoqubits} \\
    & \la_{3,\al,\be}= \left\{ -\Ga,  -\Ga  + 2\frac { \ga_{+}} {\Ga},
      -\Ga  + \frac { \ga_{+}} {\Ga} \pm 2 \De i, \right\},
      \nonumber
  \end{align}
  where $\ga_{\pm}= 4(1\pm \ga^2)$.  The respective eigenfunctions are
  fully analytic functions of $\Ga$ in the Zeno regime
  ($\Ga > \Ga_\mathrm{cr}$, see later for its definition) so the
  Liouvillian is diagonalizable in any point. In the following
  considerations, the free fermion point $\De=0$ must be excluded,
  since it corresponds to zero $h_D$ and multiple degeneracies even in
  the Zeno limit (\ref{la12-twoqubits}).

  As discussed in the main text, the analyticity of Liouvillian
  eigenvalues breaks down at the branch points, which can be located
  by finding the eigenvalues of the Liouvillian for arbitrary
  $\Ga,\ga,\De$. An inspection shows that among the 16 eigenvalues for
  $\De\neq 0$, apart from $\la=0$ there is a double degenerate real
  eigenvalue $\la= - \Ga/2$, the eigenvalue $\la=-\Ga$ and all the
  other eigenvalues contain branch points. Depending on the
  parameters, there can be up to $8$ values of $\Ga=\Ga_i$ where
  branchings occur.  Two points are $\Ga_1=8$ and $\Ga_2=8|\ga|$,
  while the location of the other branch points $\Ga_3,\dots, \Ga_8$
  involves radicals of a quartic equation.  In particular, for small
  $\De$ we find a singularity, for
  $\max(\Ga_3,\ldots \Ga_8) = O(1/|\Delta|)$, which has a probable
  origin in the repulsion of the eigenvalues, which, for
  $|\De| \ll 1$, become too close each other.  The onset of the fully
  analytic Zeno regime sets in beyond the rightmost branching points,
  i.e., for $\Ga > \Ga_\mathrm{cr} \equiv \max_i {\Ga_i}$. The value
  of $\Ga_\mathrm{cr}$ is easily estimated numerically for a generic
  choice of the model parameters, see Fig.~\ref{Fig-bifurcations} for
  an example.

%
%

\end{document}